\documentclass[aps,prl,twocolumn,superscriptaddress,showpacs]{revtex4}
\usepackage{epsfig}
\usepackage{amsmath}
\usepackage{times}

\usepackage{graphicx}
\usepackage{dcolumn}
\usepackage{bm}

\begin{document}

\title{Epidemic reemergence in adaptive complex networks}
\author{J. Zhou}
\affiliation{Division of Communication Engineering, School of
Electrical and Electronic Engineering, Nanyang Technological
University, Singapore 639798}

\author{G. Xiao}
\affiliation{Division of Communication Engineering, School of
Electrical and Electronic Engineering, Nanyang Technological
University, Singapore 639798}

\author{S. A. Cheong}
\affiliation{Division of Physics and Applied Physics, School of
Physical and Mathematical Sciences, Nanyang Technological
University, 21 Nanyang Links, Singapore 637371}

\author{X. Fu}
\affiliation{Computing Science Department, Institute of High Performance
Computing, Singapore 138623}

\author{L. Wong}
\affiliation{School of Computing \& School of Medicine, National
University of Singapore, Singapore 117417}

\author{S. Ma}
\affiliation{Epidemiology \& Disease Control Division, Ministry of
Health, Singapore 169854}

\author{T. H. Cheng}
\affiliation{Division of Communication Engineering, School of
Electrical and Electronic Engineering, Nanyang Technological
University, Singapore 639798}


\begin{abstract}
The dynamic nature of system gives rise to dynamical features of
epidemic spreading, such as oscillation and bistability. In this
paper, by studying the epidemic spreading in growing networks, in
which susceptible nodes may adaptively break the connections with
infected ones yet avoid getting isolated, we reveal a new
phenomenon - \emph{epidemic reemergence}, where the number of
infected nodes is incubated at a low level for a long time and
then bursts up for a short time. The process may repeat several
times before the infection finally vanishes. Simulation results
show that all the three factors, namely the network growth, the
connection breaking and the isolation avoidance, are necessary for
epidemic reemergence to happen. We present a simple theoretical
analysis to explain the process of reemergence in detail. Our
study may offer some useful insights helping explain the
phenomenon of repeated epidemic explosions.
\end{abstract}


\maketitle

\section{I. Introduction}
Infectious diseases have caused tremendous losses in human health
and lives and they remain as a serious threat to mankind today. To
resist infectious disease, theoretical investigations have been
engaged to study the epidemic behaviors
\cite{Bailey:1975,Satorras:2001,Anderson:1991,Barthelemy:2004,Boccaletti:2006}
and immunization strategies were suggested to prevent epidemic
spreading when vaccine resources are limited
\cite{Dezso:2002,Chen:2008,Shaw:2010,Salathe:2010}. Recently,
large-scale agent-based simulations have been applied to get more
detailed descriptions of disease outbreaks
\cite{Germann:2006,Eubank:2004,Ferguson:2005}. A prominent
development among these studies was to abstract the complex social
relations into networks, where nodes represent individuals and
links represent the contacts among them. It is found that the
basic reproductive number $R_0$, a key factor determining whether
a disease can spread out or not, depends strongly on the variance
of the distribution of the contacts
\cite{Satorras:2001,Anderson:1991,Lloyd:2001}. Extensive results
show that the social contacts typically have a fat-tail degree
distribution where a small number of people have very large number
of contacts \cite{Barabasi:1999,Liljeros:2001}. This property
typically leads to a much higher $R_0$ than that in a network with
the same average degree but homogeneous degree distribution
\cite{Satorras:2001}.

The non-trivial features of social networks such as small-world
property and fat-tail degree distribution, and the complexity of
the dynamics of infectious diseases, lead to some interesting
properties of epidemic spreading. For example, it is found that
for a linearly growing network, the evolution of the number of the
infected nodes has oscillatory behaviors when the
susceptible-infected-recovered (SIR) model is adopted
\cite{Hayashi:2004}. An adaptive mechanism is studied in
\cite{Gross:2006}, where a susceptible individual may avoid the
contacts with his infected neighbors and rewire these contacts to
other susceptible individuals. An important observation is that
the interplay of the epidemic dynamics and the network topology
may cause a bi-stable phenomenon. That is, the disease is hard to
persist when the infection density is low, yet may enter endemic
state when the infection density is high.
Network growth is a fundamental property for
any healthy systems \cite{Szell:2010} and adaptive changes widely
exist in most systems when in face of infection. Adaptations to
avoid an undesirable outcome can sometimes postpone the onset of
the undesirable outcome, at the same time making it more severe
\cite{Araujo:2010,Araujo:2011}. Hence it is necessary to study the
influences of both network growth and adaptive dynamics to the
dynamics of epidemic spreading.

In this paper, we study the epidemic spreading in linearly growing
networks assuming that the susceptible nodes may break the
contacts with the infected nodes. Considering the fact that in
general an individual cannot survive when s/he is fully isolated
in a modern society, the contact breaking process takes place only
when both of the two end nodes of a contact still have other
neighbors. Interestingly, we observe an epidemic reemergence
phenomenon, where the number of infected nodes may stay at a low
level for a long time and then bursts up to a high level. The
process may repeat for a long time before the disease finally dies
out. In Sec.~II we present the epidemic model. Simulation results
are presented in Sec.~III. In Sec.~IV we give some theoretical
analysis to explain our observations. Sec.~V presents some further
discussion and concludes the paper.

\section{II. Model}

Consider a Barabasi-Albert (BA) model \cite{Barabasi:1999} with
$N$ nodes as the initial network, where the average nodal degree
$\langle k\rangle=2m$ and $m$ is the number of links attached by each
newly added node. We use the SIS model to describe the
epidemiological process, which is widely adopted to describe
infectious diseases
\cite{Bailey:1975,Araujo:2010,Weiss:1971,Murray:1993}. In this
model agents can be in two distinct states: susceptible or
infected. A susceptible agent may get infected if he has infected
neighbor(s). Suppose a susceptible node has neighbors, of which
$k_\mathrm{inf}$ are infected, and the probability of having
contagion with each infected neighbor is $p$. Then the susceptible
node may get infected with probability $1-(1-p)^{k_\mathrm{inf}}$
which is approximately $pk_\mathrm{inf}$ when $p$ is small. At the
same time, each infected agent will become susceptible at a rate
of $r$. In this paper, we set $r=0.04$ unless otherwise specified.

\section{III. Simulation results}

\begin{figure}
\epsfig{figure=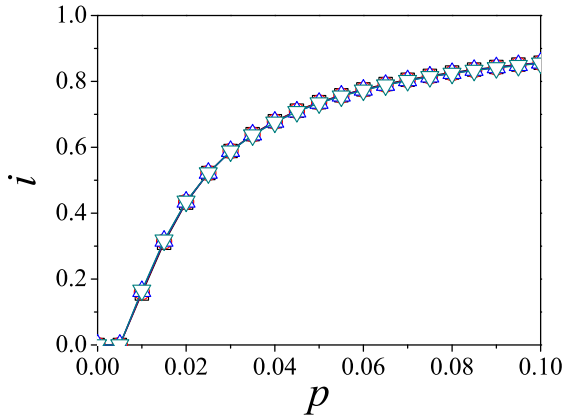,width=1\linewidth} \caption{(Color
online) The fraction of infected nodes after transient time as a
function of parameter $p$ for different growing rates
$\alpha=0.02$, $0.05$, $0.1$ and $0.2$, where $r=0.04$ and $m=2$.}
\label{Fig:Growing}
\end{figure}

We first consider the case in which the network is continually
growing during epidemic spreading. Assume on average there are
$\alpha$ susceptible nodes joining the networks at each time step
and each new node brings $m$ links connected to the existing nodes
in a preferential attachment manner. We assume that the newly
added nodes know the infectious states of the existing nodes. Thus
they only connect to susceptible ones. The probability that a
susceptible node $e$ is connected to a newly added node is
$mk_e^S/\sum_{e'}k_{e'}^S$, where $k_e^S$ denotes the degree of
the susceptible node $e$. Such a network growth model has also
been adopted in \cite{Guerra:2010}. Figure \ref{Fig:Growing} shows
the relation between the fraction of infected nodes $i$ after
transient process and the disease transmission probability $p$ for
different values of $\alpha$. In this figure, all the curves
overlap completely. Therefore, the fraction of infected nodes will
not be influenced when we change the network growing rate.

\begin{figure}
\epsfig{figure=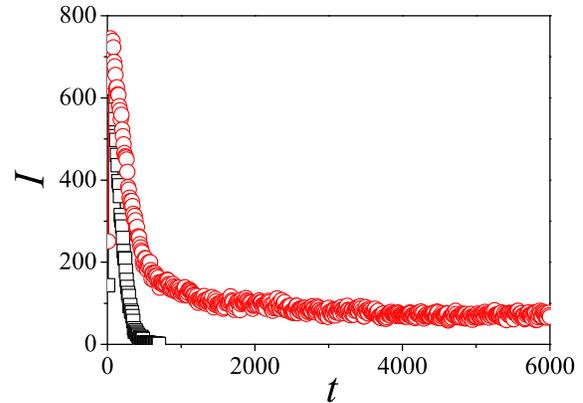,width=1\linewidth} \caption{(Color
online) The evolution of the number of infected nodes in the BA
model without network growth. We set $p=0.09$ and all the other
parameters are the same as those in figure \ref{Fig:Growing}. The
red hollow circles (black hollow squares) represent the results
when the isolation avoidance is (is not) considered. }
\label{Fig:Breaking}
\end{figure}

Now we consider the case that susceptible nodes may remove their
links connected to infected nodes when the network is not growing.
Specifically, in each time step each susceptible node may break
the link connected to an infected neighbor at a probability
$\omega$. Considering that an individual seldom can go isolation
in a modern society, we set the constraint that a link removal can
happen only when both of the two end nodes of the link have other
neighbors. We term such a constraint as \emph{isolation
avoidance}. Obviously, without isolation avoidance, the link
removal process can finally lead the network to a disease-free
status, since it is equivalent to reducing the effective infection
rate \cite{Gross:2006}, in the extreme case to zero when all the
links to the infected nodes are removed. With the isolation
avoidance, however, the link removal process cannot guarantee the
clearance of the disease, since the mechanism lets each node
retain at least one connection and a disease may spread out in the
condition of $p>r$ as the reproductive number $R_0>1$. Figure
\ref{Fig:Breaking} shows the evolution of the number of infected
nodes for the cases with and without isolation avoidance,
respectively. The infectious disease disappears after a transient
process when the isolation avoidance mechanism is not considered.
However, when the isolation avoidance is adopted, the number of
infected nodes may remain at a low level rather than disappear.

\begin{figure}
\epsfig{figure=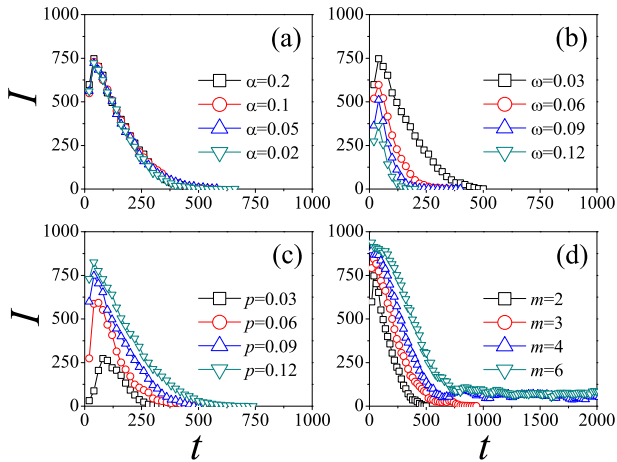,width=1\linewidth} \caption{(Color
online) Evolution of the number of infected nodes in network with
growth and link removal but without isolation avoidance. (a)
Results when $\alpha=0.2$, $0.1$, $0.05$ and $0.02$ for $m=2$,
$\omega=0.03$, $p=0.09$. (b) Results when $\omega=0.03$, $0.06$,
$0.09$ and $0.12$ for $m=2$, $p=0.09$, $\alpha=0.2$. (c) Results
when $p=0.03$, $0.06$, $0.09$ and $0.12$ for $m=2$, $\alpha=0.2$,
$\omega=0.03$. (d) Results when $m=2$, $3$, $4$ and $6$ for
$\alpha=0.2$, $\omega=0.03$, $p=0.09$.}
\label{Fig:Growing&Breaking}
\end{figure}

Next we consider the case in a growing network where susceptible
nodes may remove their links connected to infected nodes without
isolation avoidance. Figure \ref{Fig:Growing&Breaking} shows the
evolution of the number of infected nodes, $I$, with different
parameters. It shows that network growing rate $\alpha$ does not
have significant influence to epidemic dynamics; and in all the
cases, the infection will finally either go extinction or stay at
a low level.

\begin{figure}
\epsfig{figure=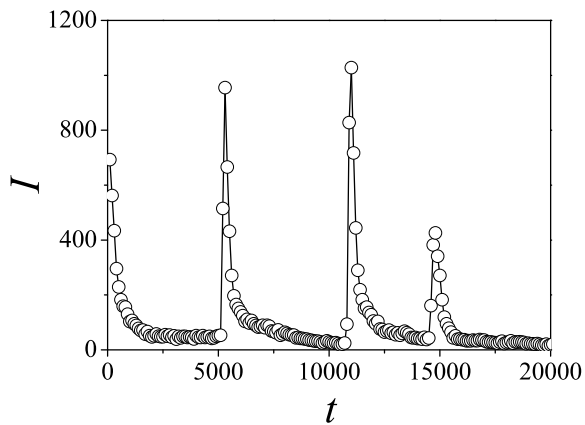,width=1\linewidth} \caption{The
evolution of the number of infected nodes when the network growth,
the link removal process and isolation avoidance are all
considered, $\alpha=0.2$, $\omega=0.03$, $p=0.09$ and $m=2$.}
\label{Fig:Growing&Breaking&IsolationAvoidance}
\end{figure}

Finally we study the case in which network growth, link removal
process and the mechanism of isolation avoidance are all involved.
Interestingly, we observe the phenomenon of epidemic reemergence.
As shown in figure \ref{Fig:Growing&Breaking&IsolationAvoidance}
where $\alpha=0.2$, $\omega=0.03$, $p=0.09$ and $m=2$, the number
of infected nodes stays at a low level for a long time, then
suddenly bursts up to a high level before decays to a low level
again. This process may repeat a certain times before the
infection finally dies out.

This observation can be explained as follows: on one hand, the
link removal process can suppress the epidemic spreading, while
the isolation avoidance lets each node have at least one neighbor.
The interplay between these two processes makes the number of
infected nodes remain at a low level. On the other hand, the newly
added nodes connect to susceptible nodes when they join the
network. Due to the small number of infected nodes and the
isolation avoidance, the infection cannot easily reach these newly
added nodes. Therefore, the newly added nodes can cumulate,
connecting susceptible nodes into a large component. Once the
infection invades into this component, however, the infection size
can quickly burst up.

\begin{figure}
\epsfig{figure=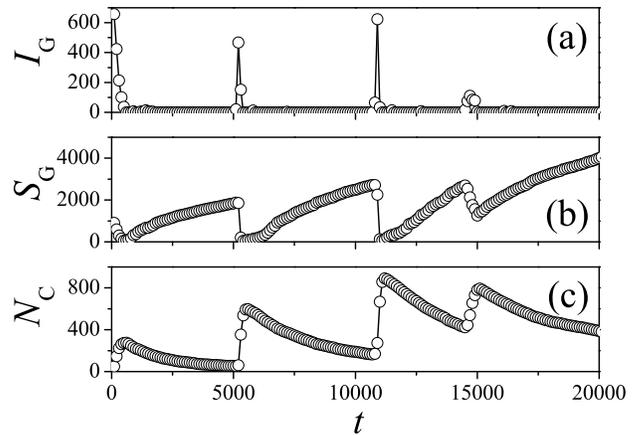,width=1\linewidth} \caption{Evolution
of the number of infected nodes in the giant component, $I_G$, the
size of giant component, $S_G$, and the number of components,
$N_C$. Parameters are the same as those in figure
\ref{Fig:Growing&Breaking&IsolationAvoidance}.}
\label{Fig:DetailedEvolution}
\end{figure}

To reveal the mechanism of the epidemic reemergence in more
detail, we illustrate the evolution of the number of infected
nodes in the giant component, $I_G$, the size of giant component,
$S_G$, and the number of components, $N_C$, respectively, in
figure \ref{Fig:DetailedEvolution}. Refer to figure
\ref{Fig:Growing&Breaking&IsolationAvoidance}, we can see that
when the number of the infected nodes $I$ is very small, the size
of the giant component increases linearly along time and $I_G$ is
about zero. This means that the giant component is basically in
the disease-free status. Meanwhile, $N_C$ decreases along time,
meaning that the newly added nodes continually merge the existing
components into larger ones. However, once the disease invades
into the giant component, the infection may quickly spread over the
whole component, leading to a high value of $I_G$. Then the
component quickly breaks into small pieces due to the link
removals, evidenced by an increasing number of $N_C$.

The growth of a large giant component plays an important role in
inducing the reemergence phenomenon. For a component that is
totally composed of susceptible nodes, referred as S-component,
the size of it may keep grow during the network growing process.
However, if a newly added node connects the S-component to a
component containing infected nodes, referred as I-component, the
infection may reach the S-component by going through the newly
added node. If the size of the S-component is small, it may not
make a large impact when it is infected. When a large S-component
is infected, however, the disease may quickly spread over it,
causing a sharp increase in the number of infected nodes.

It is interesting to have a closer look at how the small
I-components survive the long inter-epidemic periods and how the
infection finally invades a large S-component (in most cases, the
giant component). We take the second explosion which happens at
around $t=5000$ as an example. At $t=5034$, there are totally $46$
infected nodes left in the network and they belong to $5$
I-components. One of these I-components is an 18-node star network
which is composed of $11$ infected nodes including the hub of the
star and $7$ susceptible nodes. This explains how the infection
remains endemic: the links in the I-component cannot be removed
due to isolation avoidance. At $t=5035$, the 18-node I-component
is connected to the giant component through one of its susceptible
nodes. A few time steps later the infection invades the giant
component, leading to an epidemic. In fact, similar observations
apply to all the other epidemics: the long-lived I-components
seeding the epidemics almost always have star or star-like
topologies. The isolation avoidance mechanism prevents such
I-components from being further fragmented. The infection
therefore has a chance to survive over a long time. Finally, by
chance the I-component may be connected to the giant/big
S-component, which may lead to an explosion.

\begin{figure}
\epsfig{figure=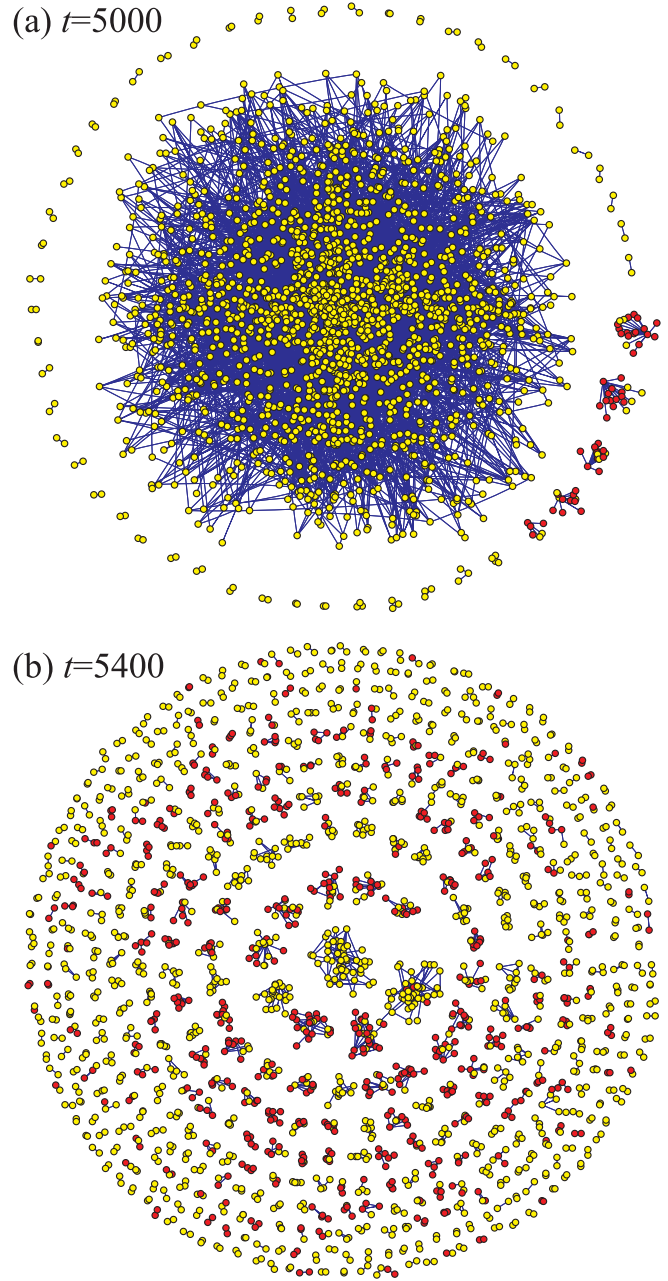,width=1\linewidth} \caption{Snapshots
of the network topology when (a) $t=5000$ and (b) $t=5400$. The
yellow (red) circles denote the susceptible (infected) nodes.}
\label{Fig:Snapshot}
\end{figure}

Figure \ref{Fig:Snapshot} illustrates snapshots of network
structures right before and after an epidemic explosion.
Specifically, when $t=5000$ which is just before the  epidemic
explosion, as we can see in Figure \ref{Fig:Snapshot}(a), there is
a giant component containing most of the network nodes and all the
nodes in this component are susceptible. Meanwhile, a small number
of infected nodes exist in several small components. This
observation is accordant with our previous discussion that due to
the network growth and isolation avoidance, the network can form a
giant component which is basically disease free, while a small
number of infected nodes may remain in small clusters. Figure
\ref{Fig:Snapshot}(b) shows the network structure when $t=5400$
which is right after the epidemic explosion. We see that the
network breaks into many small pieces, most of which with no more
than $5$ nodes. In this stage, on one hand, the network still has
many infected nodes, thus some I-components may be further
decomposed to suppress the infection spreading; on the other hand,
the network has several S-components with relatively large size,
which indicates that the new giant component starts to form.

Note that a reconnection between an I-component and the giant/big
S-component does {\it not} always lead to an explosion. In fact, in most
cases, the giant/big component is finally disconnected from any
infected nodes again without any noticeable increase in the number
of infected nodes in between. An explosion typically only happens
when the infection manages to reach a high-degree hub in the giant
component. In the next section, we will show in more detail the
strong randomness in the reconnection process.

\section{IV. Theoretical analysis}

Now we do some theoretical analysis to explain the observations in
the simulation. We start with the status that the network has just
suffered from a large-scale infection and the susceptible nodes
have cut off a large number of connections to protect themselves.
Consequently, the network is broken into pieces. In such
situation, though most nodes have recovered to susceptible status,
due to the isolation avoidance, there are still a small number of
infected nodes remaining in the network. Because the number of
infected nodes and the number of the corresponding I-components
are very small, the probability that a newly added node connects
to a susceptible node belonging to an I-component is low. Hence,
to study the growing speed of the giant component, we can ignore
the effects of the infected nodes and the corresponding
I-components. Upon obtaining the growing behavior of the giant
component, we can then calculate the expected time that the giant
component is connected to an I-component.

We assume that there are $N_0$ nodes and $M_0/2$ links remaining
in a network just suffered from a large-scale infection. Therefore
on average each node has a degree $M_0/N_0$. Denote the average
component size as $\langle g\rangle$. We have $\langle
g\rangle=\sum_jg_j/N_G$ where $g_j$ denotes the size of component
$j$ and $N_G$ the number of the components. Since the probability
that a new node connects to a component is proportional to the
number of links remaining in the component and most nodes in the
small components are of very low and similar nodal degrees,
approximately the probability that a new node connects to a
component is proportional to the size of this component. Therefore
the average size of the components, other than the giant one, that
are chosen by a new node is $h=\langle g^2\rangle/\langle
g\rangle$, where $\langle g^2\rangle=\sum_jg_j^2/N_G$. Ideally, if
a new node connects to only a single node in the giant component
and connects to other $m-1$ nodes from other different components,
the size of the giant component may be increased by $(m-1)h+1$.
However, with the growing of the giant component, the probability
of having the new node making multiple connections to the giant
component increases. Thus the giant component size may be
increased at a speed of $(hl+1)\alpha$, where $l$ denotes the
number of components connected to the new node other than the
giant one, $l\in[0,m-1]$. We assume that each new node connects
the giant component by at least one link when he joins the
network. This assumption is reasonable: on one hand, when all
components are of small sizes, a new node may not connect to the
largest component at that time. However, since the new node
connects a few small components into a larger one, the newly
formed component stands a better chance to be connected to more
new nodes arriving later and become even bigger. With the growing
of the network, the newly formed component has a high chance to be
finally included into the largest component. Once this happens,
the new node eventually contributes to increasing the size of the
giant component. On the other hand, when the giant component is
large, the probability that it is connected to a new node is high.
Based on the above consideration, the evolution of the giant
component size, denoted as $S_G$, can be expressed as follows:

\begin{equation}
\label{eq:dSGdt}
\frac{dS_G}{dt'}=\alpha\sum_{l=0}^{m-1}C_{m-1}^l\left(1-\frac{M_G(t')}{M(t')}\right)^l
\left(\frac{M_G(t')}{M(t')}\right)^{m-l-1}(h\cdot
l+1)\,,\end{equation} where $t'$ is the elapsed time after the
network breaks into pieces, $M(t')=M_0+2m\alpha t'$ is the sum of
the degrees of all the nodes at time $t'$, and
$M_G(t')=(S_G-\alpha t')M_0/N_0+2m\alpha t'$ is the sum of the
degrees of all the nodes in the giant component at time $t'$. Note
that $S_G-\alpha t'$ equals the number of nodes when $t'=0$. The
term in the first (second) bracket in the summation of
Eq.~(\ref{eq:dSGdt}) indicates the probability that a link sourced
from a new node does not (does) choose the giant component, in
which the preferential attachment is adopted and the effects of
I-components are ignored. For the case of $m=2$,
Eq.~(\ref{eq:dSGdt}) goes to
\begin{eqnarray}
\label{eq:dSGdt-m=2} \frac{dS_G}{dt'}&=&\left(1-\frac{(S_G-\alpha
t')\cfrac{M_0}{N_0}+4\alpha t'}{M_0+4\alpha
t'}\right)(h+1)\alpha\nonumber\\&+&\frac{(S_G-\alpha
t')\cfrac{M_0}{N_0}+4\alpha t'}{M_0+4\alpha
t'}\alpha\,.\end{eqnarray} Solving Eq.~(\ref{eq:dSGdt-m=2}), we
have
\begin{equation}
\label{eq:SG} S_G=(N_0+\alpha t')+(M_0+4\alpha
t')^{-\frac{hM_0}{4N_0}}\left(hM_0^{\frac{hM_0}{4N_0}}-N_0M_0^{\frac{hM_0}{4N_0}}\right)\,,
\end{equation}
where we set $S_G(0)=h$.

\begin{figure}
\epsfig{figure=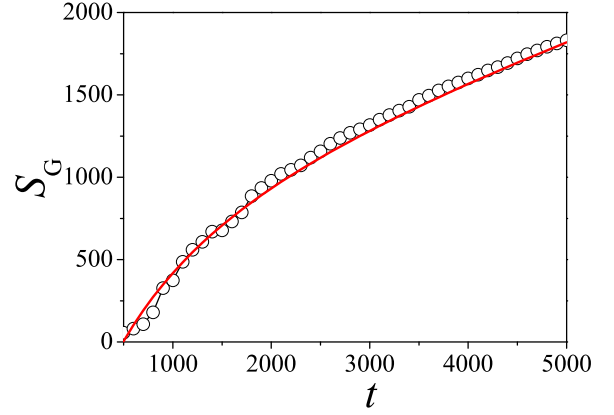,width=1\linewidth} \caption{(Color
online) A detailed plot of figure \ref{Fig:DetailedEvolution}(b)
for $t\in[500,5000]$. The circles show the simulation results and
the red solid curve shows the theoretical results obtained from
Eq. (\ref{eq:SG}). In the theoretical analysis, we adopt the
calculation results that at time $t=500$ (i.e., $t'=0$),
$N_0=1100$, $M_0=1648$, $h=4.15$. Other parameter values include
$\alpha=0.2$ and $m=2$.} \label{Fig:SG}
\end{figure}

Figure \ref{Fig:SG} shows the results of figure
\ref{Fig:DetailedEvolution}(b) in the range of $t\in[500,5000]$.
When $t$ is around $500$, the network breaks up to a large
number of small pieces. After that the network grows continually
until a giant component forms up. In this figure, the circles show
the simulation results and the red solid curve shows the
theoretical results obtained from Eq.~(\ref{eq:SG}). We can see
they match fairly well.

Following we analyze the average duration when the giant component
can keep growing without getting connected to an I-component.
Denote $P(T)$ as the probability that the giant component connects
to I-component for the first time at time $T$. We have

\begin{equation}
\label{eq:PT}
P(T)=\left(\prod_{t'=0}^{T-1}\Theta(t')^\alpha\right)\left(1-\Theta(T)^\alpha\right)\,,
\end{equation}
where
\begin{eqnarray}
\label{eq:Theta}
\Theta(t')&=&\left(\frac{M(t')-M_I}{M(t')}\right)^m+\left(\frac{M(t')-M_G(t')}{M(t')}\right)^m\nonumber\\
&-&\left(\frac{M(t')-M_I-M_G(t')}{M(t')}\right)^m\,.
\end{eqnarray}
In Eq.~(\ref{eq:Theta}), $M_I$ is the number of susceptible nodes
in the I-components which does not change a lot during the network
growing process until the infection invades the giant component,
and hence is treated as a constant value here. The first (second)
term on the right-hand side indicates the probability that none of
the $m$ links of the new node connects to the I-components (giant
component), while the third term indicates the probability that
none of the $m$ links connects to either the I-components or the
giant component. Therefore, $\Theta(t')$ presents the probability
that a node newly added at time $t'$ does not connect any
I-component to the giant component. As $\alpha$ is the growing
rate of the new nodes, $\Theta(t')^\alpha$ represents the
probability that all the newly added nodes do not connect any
I-component to the giant component at time $t'$. Hence, the
expected value of the time that the giant component gets connected
to an I-component for the first time is
\begin{eqnarray}
\label{eq:ExpectedTime}
E(T)&=&\sum_{T=1}^{\infty}T\cdot P(T)\nonumber\\
&=&\sum_{T=1}^{\infty}T\cdot\left(\prod_{t'=0}^{T-1}\Theta(t')^\alpha\right)\left(1-\Theta(T)^\alpha\right)\,,
\end{eqnarray}
which can be simplified to
\begin{equation}
\label{eq:CompactedExpectedTime}
E(T)=\sum_{T=0}^{\infty}\left(\prod_{t'=0}^{T}\Theta(t')^\alpha\right)\,.
\end{equation}

\begin{figure}
\epsfig{figure=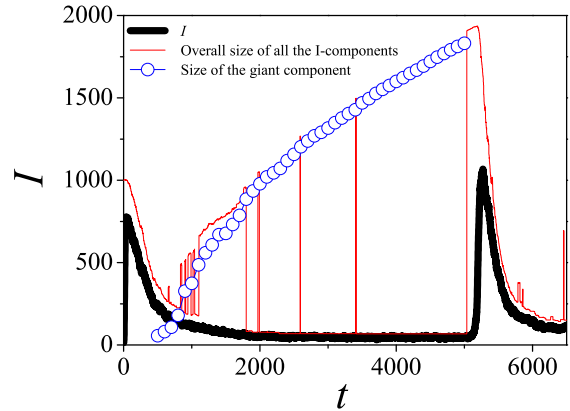,width=1\linewidth} \caption{(Color
online) The evolution of the overall size of all the I-components
indicated by the thin red curve, the number of infected nodes
indicated by the thick black curve, and the size of the giant
component indicated by the blue circles. All parameters are the
same as those in figure \ref{Fig:SG}.} \label{Fig:Jumps}
\end{figure}

Setting $M_I=20$, which is obtained from the simulation, and all
the other parameters the same as those in figure \ref{Fig:SG}, we
substitute Eq. (\ref{eq:SG}) into
Eq.~(\ref{eq:CompactedExpectedTime}) and then solve it
numerically. The result is $E(T)\simeq 723$. Using similar method,
we can calculate the standard deviation as $522$. The very large
standard deviation shows that there exist strong fluctuations in
the time intervals between reconnections of I-component and the
giant component. To demonstrate such strong fluctuations, we plot
in figure \ref{Fig:Jumps} the evolution of the overall size of all
the I-components represented by the thin red curve. It is easy to
understand that the overall infection size would jump up when a
small I-component is connected to the giant component (which makes
the giant component itself an I-component, evidenced by the
heights of the jumps closing to the size of the giant component).
As we can see, after the transient process, several jumps take
place at around time $1970$, $2590$, $3405$ and $5035$ and the
time intervals in between are $620$, $815$ and $1630$,
respectively. Such results are accordant with our theoretical
estimation on the expected value of time interval and its
deviation. The observation on the large deviation value has its
significance: the strong fluctuations of the reconnection process
make it difficult to predict when the next reconnection of an
I-component and the giant component would happen though we could
estimate the long-term average of the intervals. Further consider
the fact that a reconnection (reflected as a peak of the thin red
curve in figure \ref{Fig:Jumps}) does not always lead to an
explosion (a peak of thick black curve in figure \ref{Fig:Jumps}),
it would be very difficult, if not impossible, to predict when a
reemergence would happen.

Once the infection does invade the giant component, the
disease may spread over the whole component very quickly and
further transmit to other smaller components. In fact, in the
beginning stage of the outbreak, the fraction of infected nodes
grows exponentially fast \cite{Barthelemy:2004}. Specifically,
with mean-field approximation the evolution of $i$ and $\langle
k\rangle$ can be expressed as

\begin{equation}
\label{eq:didt&dkdt} \left\{
\begin{array}{ccc}
\cfrac{di}{dt}=p\langle k\rangle (1-i)i-ri \\\\
\cfrac{d\langle k\rangle}{dt}=-2\omega\langle k\rangle (1-i)i
\end{array}
\right.\,.
\end{equation}
Here we ignore the network growing process since we mainly focus
on the drastic change of the infection size in which the time span
is short and the number of newly added nodes is very limited.
Since $i$ is very small in the beginning stage of epidemic
explosion, we neglect the high-order terms of $i$. Besides,
because of the small time interval during this process, we regard
$\langle k\rangle$ as a constant. Thus, the first equation in Eq.
(\ref{eq:didt&dkdt}) can be simplified as \cite{Barthelemy:2004}
\begin{equation}
\label{eq:didt-simplified} \frac{di}{dt}=p\langle k\rangle i-ri\,.
\end{equation}
Solving Eq.~(\ref{eq:didt-simplified}), we have
\begin{equation}
\label{eq:i-simplified} i(t)=i(0)e^{(p\langle k\rangle -r)t}\,.
\end{equation}
We see that $i(t)$ grows exponentially, which proves that the
number of infected can increase very quickly in a short time.

Figure \ref{Fig:Growing&Breaking&IsolationAvoidance} illustrates
that after each infection peak, the number of infected nodes
decreases gradually and this decreasing process sustains much
longer than the increasing process. To study the behaviors of the
decreasing process, we first consider the dynamics of $i(t)$
during a short period of time where $\langle k\rangle$ can be
approximately regarded as a constant. Solving the first equation
in Eq.~(\ref{eq:didt&dkdt}) by assuming $\langle k\rangle$ as a
constant, we have
\begin{equation}
\label{eq:i-kFixed} i(t)=(p\langle
k\rangle-r)\left(1-\frac{1}{1+Cp\langle k\rangle e^{(p\langle
k\rangle -r)t}}\right)\,,
\end{equation}
where $C=\cfrac{i(0)}{p\langle k\rangle(1-i(0))-r}$.

\begin{figure}
\epsfig{figure=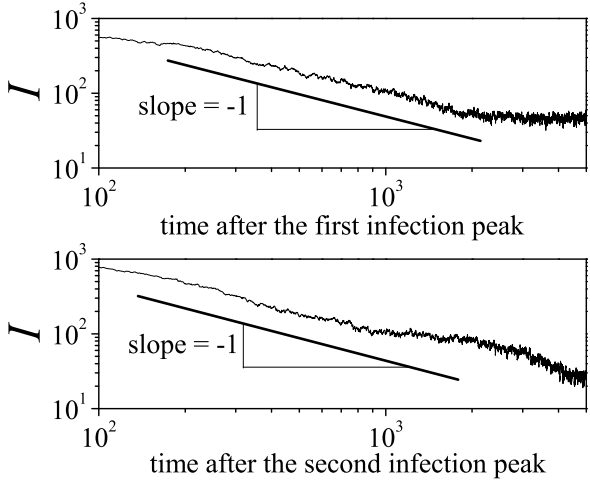,width=1\linewidth} \caption{The
evolutions of the number of infected nodes after the first and
second infection peak. Parameters are the same as those in figure
\ref{Fig:SG}.} \label{Fig:slope}
\end{figure}

The exponential function in Eq.~(\ref{eq:i-kFixed}) reveals that
the value of $p\langle k\rangle-r$ determines the typical time it
takes for $i(t)$ to go to the stable state for a constant value
$\langle k\rangle$. A larger $p\langle k\rangle-r$ corresponds to
a shorter time. For the specific case as shown in figures
\ref{Fig:Growing&Breaking&IsolationAvoidance} and \ref{Fig:Jumps}
where $m=2$, $p=0.09$, $r=0.04$ and $\omega=0.03$, when $i(t)$
starts to decrease, $\langle k\rangle\approx 3$ and $(p\langle
k\rangle-r)/\omega\approx 8$. This shows that the spreading
process is much faster than link removal process. Such observation
allows us to simplify our analysis by considering the decreasing
process as composed of a series of quasi-static processes. That
is, we divide the decreasing process into a series of very short
time intervals and regard each short interval as composed of two
different parts: first some links are removed; then $i(t)$ quickly
converges to a temporally stable state where $di/dt=p\langle
k\rangle (1-i)i-ri=0$ for the updated value of $\langle k\rangle$.
Hence, suppose $\langle k\rangle$ has a small change in each time
interval as $\langle k\rangle\rightarrow\langle
k\rangle-\varepsilon$, where $\varepsilon$ is a small positive
quantity. We have
\begin{equation}
\label{eq:didt-QuasiStatic} \frac{di}{dt}=p(\langle
k\rangle-\varepsilon)(1-i)i-ri=-\varepsilon p(1-i)i\,.
\end{equation}
By solving Eq.~(\ref{eq:didt-QuasiStatic}), the evolution of
$i(t)$ can be expressed as $i(t)=1/(C'e^{\varepsilon pt}+1)$,
where $C'$ is a constant depending on the initial condition. As
$\varepsilon$ is of a very small positive value, $i(t)$ can be
further approximated as
\begin{equation}
\label{eq:i-QuasiStatic} i(t)=\frac{1}{1+C'\varepsilon pt+C'}\sim
\frac{1}{t}\,.
\end{equation}
Therefore, we have that approximately $i(t)$ evolves
inverse-proportionally with time $t$ after the infection peak. The
log-log plot in figure \ref{Fig:slope} shows the simulation
results of the evolution of the number of infected nodes $I$ after
the first and second infection peaks for the case in figure
\ref{Fig:Jumps}, where the first (second) peak happens around the
time $t=100$ ($t=5270$). As expected, $I$ evolves approximately
inverse proportionally with time when it decays from the peak.

When the number of infected nodes decreases to a small value, it
will become stable. In this stage, the network breaks to an extent
that the sizes of I-components are small enough so that the
breaking process almost ceases due to the isolation avoidance.
Consequently a small number of infected nodes may be preserved in
the I-components for a long time until the next epidemic explosion
happens.

\begin{figure}
\epsfig{figure=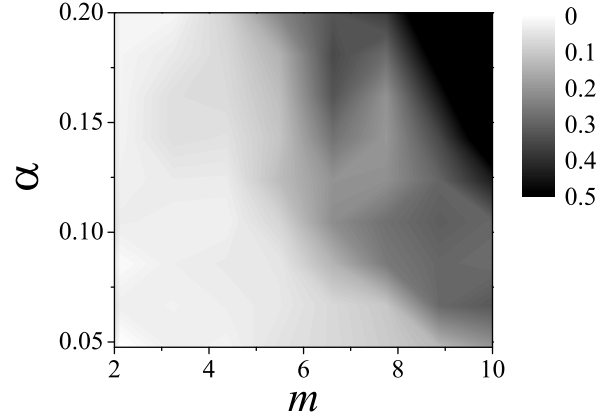,width=1\linewidth} \caption{Gray plot
of parameter $F$ on the $m-\alpha$ plane. Results are obtained
from $100$ realizations. Other parameters are the same as those in
figure \ref{Fig:Growing&Breaking&IsolationAvoidance}.}
\label{Fig:GrayPlot}
\end{figure}

It has been observed that epidemic reemergence may happen for a
wide range of $\alpha$ and $m$. To manifest the influences of
$\alpha$ and $m$, we introduce a parameter $F$ to measure the
burstiness of the epidemic reemergence, which is defined as
\begin{equation}
\label{eq:F}
F=\frac{T\left(I>\cfrac{I_\mathrm{max}}{2}\right)}{T\left(I>0\right)}\,,
\end{equation}
where $I_\mathrm{max}$ is the maximum value of $I$ in the whole
disease spreading process. A smaller value of $F$ corresponds to a
more abrupt burst in the number of infected nodes. Figure
\ref{Fig:GrayPlot} shows the results of parameter $F$ on the
$m-\alpha$ plane. It can be seen that generally speaking slower
network growing speed $\alpha$ and smaller $m$ lead to smaller $F$
value. This phenomenon can be understood as follows. Having a
higher network growing speed and more links added to the network
by each newly added node makes the S-components connect to
I-components in a larger probability; consequently the susceptible
nodes may get infected more easily. If a significant number of
S-components are invaded by infection before they form into a
large component, the reemergence shall become less bursty.
Note that in all our
simulations, the infection finally dies out: small fluctuations at
a very low infection size sooner or later will push the infection
into extinction. Reemergence cannot be repeated forever unless it
has a stable pool outside the system we have been considering.

\section{V. Discussion and conclusion}

Based on a simple model which captures basic characteristics of
epidemic spreading, we revealed the mechanism of a new phenomenon
- \emph{epidemic reemergence}. Specifically, we considered the
epidemic spreading in a growing network where a susceptible node
may remove the link connecting to an infected neighbor. Such
link-removal process, however, is subject to the isolation
avoidance constraint. It is observed that epidemic reemergence may
happen in such a simple model: the number of infected nodes may be
suppressed to a low level for a long time and then suddenly bursts
into a high level. Our findings indicate that an infectious
disease may break out repeatedly, and between two explosions the
disease may be incubated for a long time by a small number of
carriers who are not connected to the main body of the society but
are confined in a small area. The small number of disease carriers
however stands a high chance to induce a large-scale epidemic
explosion even when most people are intentionally avoiding the
disease.

In the current model, SI links are simply removed unless
the removal will cause isolation. In real life, however, people
may tend to enhance existing SS links or even build up new ones to
make compensations to the social connections they have lost. For
example, they may tend to spend more time with their family
members while reducing social gatherings. How such "rewiring"
operations and connection enhancements will affect the epidemic
spreading and reemergence is of our future research interest.

\acknowledgments

This work was supported in part by A*STAR BMRC Program under grant
contract R-252-000-297-305 and Ministry of Education, Singapore,
under contract 27/09.

\end{document}